\documentclass[reprint,superscriptaddress,amsmath,amssymb,aps,pre]{revtex4-2}

\usepackage{graphicx}
\usepackage{booktabs}
\usepackage{subcaption}
\usepackage{bm}
\usepackage{amsmath,amssymb}
\usepackage[ruled,linesnumbered]{algorithm2e}
\usepackage{booktabs}

\usepackage[table]{xcolor}

\definecolor{maxphi}{RGB}{180,220,180}

\definecolor{minphi}{RGB}{245,200,200}

\begin{document}

\title{Saturation Coverage in Binary Mixtures of Oriented Regular Polygons via Random Sequential Adsorption}

\author{Aref Abbasi Moud}
\email{abbasimoudaref@gmail.com}

\begin{abstract}
We study saturation in two-dimensional binary mixtures of fixed-orientation
regular polygons deposited via random sequential adsorption (RSA). Polygons
with $n \in \{3,\dots,23\}$ are considered under an equal-area constraint to
isolate shape effects from size effects. Saturated configurations are generated
using an adaptive split-voxel RSA algorithm with exact overlap detection via
the Separating Axis Theorem, enabling a representative mapping of all unique binary
shape combinations.
 
A strong dependence of the jamming coverage on polygon geometry is observed
despite identical particle area. Triangle-containing mixtures systematically
suppress saturation, while axis-aligned squares achieve the maximum measured
coverage $\phi_{\rm sat} \approx 0.5646$. Even-sided polygons yield
systematically higher coverage than their odd-sided neighbours, revealing a
parity-driven adsorption advantage rooted in centrosymmetry. For the
odd-$n$ sub-sequence ($n=5,7,9,11,13,17$), the pure-species saturation converges to
the classical disk RSA limit $\phi_{\rm disk}\approx 0.547$ from below,
following $\phi_{\rm sat}(n)=\phi_{\rm disk}-c/n^{\alpha}$ with fitted
exponent $\alpha\approx 2.41\pm0.06$, close to the $1/n^{2}$ scaling
predicted by isoperimetric arguments. Even-sided polygons instead approach
$\phi_{\rm disk}$ from above, confirming that centrosymmetric shapes retain
a tiling advantage that vanishes only in the circular limit.
 
These trends are explained by the excluded area
$E_{AB}=\mathrm{Area}(P_A\oplus(-P_B))$, computed analytically via
Minkowski sums. Centrosymmetry fixes $E_{AA}=4A_0$ for even $n$, whereas
odd $n$ yields the larger value
$E_{AA}=4A_0\!\left[1+\tfrac{1}{2}(\sec(\pi/n)-1)\right]$, which
decreases monotonically to $4A_0$ as $n\to\infty$. Saturation coverage is
negatively correlated with $E_{AB}$, consistent with a mean-field RSA
description based on an effective excluded area, directly linking geometric
symmetry to jamming efficiency. 
\end{abstract}

\maketitle

\section{Introduction}
 
Random sequential adsorption (RSA) is a model of irreversible deposition in
which particles are placed sequentially at random positions and remain fixed
upon successful insertion. Introduced by R\'enyi~\cite{Renyi1958} in the
one-dimensional car-parking problem, RSA has since become a standard framework
for studying colloidal deposition, protein adsorption, granular jamming, and
surface functionalization~\cite{Evans1993,Kubala2022}. The principal quantity
of interest is the saturation (jamming) coverage
\begin{equation}
  \phi_{\rm sat}=\lim_{t\to\infty}\frac{N(t)\,A_p}{A_{\rm domain}},
\end{equation}
which measures the fraction of the domain area occupied by particles in the
fully jammed state.
 
RSA has been studied for a wide variety of particle geometries \cite{Tarjus1991,zhang2018precise,Talbot1989,Viot1992,Adamczyk1997} The saturation
coverage for disks,
\begin{equation}
  \phi_{\rm disk}\approx 0.547,
\end{equation}
is a classical benchmark~\cite{Feder1980}. Subsequent work extended RSA to
rectangles~\cite{Hinrichsen1986}, ellipses, and convex
polygons~\cite{Tarjus1991,zhang2018precise}, establishing that particle shape
strongly influences jamming even when particle size is held fixed. For
fixed-orientation regular polygons specifically, Cie\'sla \textit{et al.}
identified a parity-dependent transition in packing-growth
kinetics~\cite{Ciesla2023}, and earlier single-species studies showed that
$\phi_{\rm sat}$ depends sensitively on both polygon order and
orientation~\cite{Zhang2013}. RSA is also widely used as a practical tool for
modelling macromolecule and colloidal particle deposition, where particle shape
and competitive adsorption directly control monolayer density and
microstructure~\cite{Kubala2022}.
 
Despite this body of work, binary RSA mixtures in which \emph{shape alone}
governs the competition between species have received comparatively little
attention. Studies of binary systems have largely focused on size
polydispersity~\cite{Hassan2002,Kundu2022} or mixed disk--discorectangle
systems~\cite{Lebovka2024}, leaving shape-driven competition in equal-area
mixtures unexplored. Understanding such systems is of practical relevance: in
competitive adsorption from solutions containing molecules of different shapes
but similar sizes, it is the excluded-area geometry rather than steric bulk
that determines which species preferentially occupies the surface \cite{Adamczyk1996}.
 
We investigate random sequential adsorption of equal-area regular polygon
mixtures to isolate the influence of particle geometry from size effects.
Using an adaptive split-voxel RSA algorithm combined with exact overlap
testing based on the Separating Axis Theorem~\cite{de2000computational},
we examine all 52 distinct binary combinations of regular polygons with
$n\in\{3,\ldots,23\}$ sides. This exhaustive dataset enables the
construction of a saturation-coverage landscape spanning the full range of
shape pairings and provides a direct connection between jamming behavior
and excluded-area geometry derived from Minkowski-sum analysis.

The results reveal that shape alone strongly influences both the overall
saturation coverage and the partitioning of occupied area between competing
species. In the single-species limit, the saturated coverage approaches the
disk RSA value according to the isoperimetric-deficit scaling
$\mathcal{D}(n)\propto n^{-2}$. Moreover, a systematic parity effect
emerges: even-sided polygons converge to the disk limit from above,
whereas odd-sided polygons converge from below, reflecting the role of
centrosymmetry in determining excluded-area properties. These findings are
relevant to competitive adsorption processes in colloidal and biological
systems, where particles of comparable size but differing shape compete
for finite surface resources~\cite{Ramsden1993}.

\section{Model and Methodology}

 RSA is a stochastic irreversible deposition process in which trial
particles are placed at uniformly distributed random positions over the
domain and accepted only if they do not overlap any previously deposited
particle; rejected trials are discarded and the process continues until
no further insertions are possible~\cite{Evans1993,Feder1980}. The
fraction of the domain covered at this jammed state defines the
saturation coverage $\phi_{\rm sat}$.

In order to improve computational efficiency, the configuration space
is discretized into small candidate-centre regions (voxels), which are
dynamically tested for the feasibility of particle insertion~\cite{zhang2018precise}.
Voxels that cannot accommodate the centre of a new particle without
violating the non-overlap constraint are removed from the admissible
set, thereby restricting sampling from the full domain to a reduced
effective region. The system is considered saturated when no admissible
voxels remain, at which point exact jamming is certified~\cite{zhang2018precise}.

In this framework, regular polygons are characterised by a discrete
number of sides $n$, with the circular RSA limit recovered as
$n\to\infty$~\cite{Ciesla2023}. All polygons are assumed to have fixed
orientation, so that only translational degrees of freedom are
considered~\cite{Zhang2013}. This voxel-based RSA procedure is
naturally extended to binary mixtures of polygons, where two species
with different geometries are deposited sequentially while enforcing
both intra-species and cross-species exclusion rules~\cite{Hassan2002}.

\subsection{Numerical parameters}
 
All simulations were performed on a square domain of size $L \times L$ with fixed boundary conditions \cite{ciesla2018boundary}, where $L = 20$. The area of each particle was fixed at $A_0 = 0.1$ throughout all simulations. Each reported datum is the average of $N_{\rm run}$
independent realizations ($N_{\rm run}=100$ unless stated otherwise),
evolved until the voxel elimination criterion certifies saturation.

The voxel subdivision terminates when a voxel half-width falls below a
prescribed threshold $\delta_{\min}$. In the present simulations
$\delta_{\min} = 10^{-5}$ (in units of the domain side $L = 20$),
corresponding to an absolute positional resolution of
$\delta_{\min} L = 2\times 10^{-4}$. At this scale the residual
uncovered area per voxel is at most $(\delta_{\min} L)^2 = 4\times 10^{-8}$,
which is negligible relative to the particle area $A_0 = 0.1$. Voxels
smaller than $\delta_{\min}$ that cannot be proven unavailable by the
geometric tests are conservatively treated as blocked, introducing a
systematic underestimate of $\phi_{\rm sat}$ bounded by
\begin{equation}
  \Delta\phi \;\leq\; \frac{N_{\rm vox}^{\rm final}\,(\delta_{\min} L)^2}{L^2}
  \;\ll\; 10^{-4},
\end{equation}
where $N_{\rm vox}^{\rm final}$ is the number of voxels remaining at
termination. This truncation error is two orders of magnitude smaller
than the run-to-run standard deviation $\sigma \sim 10^{-3}$ and does
not affect any reported result.
 
\subsection{Particle geometry}
 
A regular $n$-gon of side length $s$ has area
\begin{equation}
  A = \frac{ns^2}{4\tan(\pi/n)}.
\end{equation}
All species are rescaled to the same target area $A=A_0$, so that
\begin{equation}
  s_\alpha = \sqrt{\frac{4A_0\tan(\pi/n_\alpha)}{n_\alpha}},
  \qquad \alpha\in\{A,B\},
\end{equation}
ensuring that any difference in saturation coverage is purely geometric
in origin~\cite{Ciesla2023}.
 
\subsection{Split-voxel RSA algorithm}
 
Conventional RSA slows severely near saturation as the admissible
insertion region fragments into tiny disconnected patches~\cite{Feder1980}.
The split-voxel method~\cite{zhang2018precise} addresses this by
representing the configuration space as a set of adaptive
candidate-center regions (voxels) and recursively refining them until
every voxel is provably unavailable, at which point exact saturation
is certified.
 
The present implementation follows Zhang~\cite{zhang2018precise} but
differs in two key respects. First, orientations are fixed, reducing
each particle's degrees of freedom from $(x,y,\theta)$ to $(x,y)$ and
the voxel space from three to two dimensions; this eliminates the
rotational envelope and substantially reduces the voxel-explosion
problem noted in~\cite{Zhang2013}. Second, two distinct species are
deposited simultaneously with equal probability $p_A=p_B=\tfrac{1}{2}$,
so a voxel is pruned only when it is unavailable to \emph{both} species:
\begin{equation}
  \mathrm{Unavail}(V)
  = \mathrm{Unavail}_A(V) \;\land\; \mathrm{Unavail}_B(V).
\end{equation}
The algorithm proceeds by (i)~sampling a trial center from a randomly
chosen voxel, (ii)~testing insertion via hierarchical overlap screening,
(iii)~removing voxels proven unavailable after each accepted particle,
and (iv)~subdividing voxels of uncertain status into four children.
This concentrates computation in geometrically constrained regions
while avoiding high rejection rates near jamming.
 
\subsection{Overlap detection}
 
Candidate pairs are first screened by inscribed and circumscribed
radii; only geometrically ambiguous configurations proceed to exact
testing. Exact overlap between two convex polygons is determined by
the Separating Axis Theorem (SAT)~\cite{Ericson2005}: the polygons are
disjoint if and only if a separating axis exists along which their
projection intervals do not overlap. For cross-species pairs the SAT
axes are taken from the union of edge normals of both polygons.

\section{Saturation Algorithm and Convergence Certificate}
\label{sec:algorithm}

\subsection{Overview}

Random sequential adsorption (RSA) reaches a \emph{jammed} or
\emph{saturated} state when no additional particle can be inserted into the
simulation domain without overlapping one already present.  Monitoring the
acceptance rate alone is insufficient to certify saturation: because
acceptance events become exponentially rare near jamming, even a very long
run leaves an acceptance-rate estimate that is positive even when the
configuration is effectively jammed.  We therefore use a unified
voxel-subdivision algorithm that produces a \emph{provable} geometric
certificate of saturation at the end of each run.

The algorithm maintains a list of square axis aligned \emph{voxels} that collectively cover all regions of the domain
where a new particle \emph{might} still be insertable.  The simulation
terminates, with saturation certified, when this list is empty: no region of
the domain remains that could possibly accommodate a new particle.

\subsection{Unified voxel algorithm}
\label{sec:alg-unified}

The valid insertion domain for species $\alpha$ in a box $[0,L]^2$ is
\begin{equation}
  \Omega_\alpha = [b_L^\alpha,\; L - b_R^\alpha]
                \times [b_B^\alpha,\; L - b_T^\alpha],
\end{equation}
where $\mathrm{BBOX}^\alpha = [b_L^\alpha, b_R^\alpha, b_B^\alpha, b_T^\alpha]$
are the actual per-direction extents of the polygon from its centre
(defined precisely in Section~\ref{sec:border}).  The algorithm is
initialised with a single voxel
\begin{equation}
  V_0 = \Bigl(\tfrac{L}{2},\;\tfrac{L}{2},\;
               \tfrac{L}{2} - \delta_0\Bigr),
\end{equation}
where the voxel is represented as (centre $x$, centre $y$, half-width
$\delta$) and $\delta_0 = \min_{\alpha,d} b_d^\alpha$ is the smallest
border clearance across both species and all four directions.  This single
voxel exactly covers the intersection $\Omega_A \cap \Omega_B$.

Each iteration of the main loop proceeds as follows.
\begin{enumerate}
  \item \textbf{Bulk cull.}  If the voxel list exceeds 500 entries, all
        voxels that can be proved unavailable (Section~\ref{sec:prove}) are
        discarded before proceeding.  This prevents unbounded list growth.

  \item \textbf{Sample.}  A voxel $V_i = (c_x, c_y, \delta)$ is chosen
        uniformly at random from the list.  A trial centre
        $(x,y) \sim \mathrm{Uniform}([c_x \pm \delta] \times [c_y \pm \delta])$
        is drawn, and a species $\alpha \in \{A, B\}$ is chosen with equal
        probability.

  \item \textbf{Accept.}  If the polygon of species $\alpha$ placed at
        $(x, y)$ does not overlap any already-deposited particle and lies
        within $\Omega_\alpha$, the trial is accepted.  The new particle is
        added to the configuration, and every voxel in the list that can now
        be proved unavailable is immediately discarded.

  \item \textbf{Reject and subdivide.}  If the trial is rejected, one of
        three outcomes follows for voxel $V_i$:
        \begin{enumerate}
          \item[(a)] \emph{Delete:} if $\delta < \delta_{\min}$, or if $V_i$
                can be proved unavailable, it is removed from the list.
          \item[(b)] \emph{Subdivide:} otherwise $V_i$ is replaced by its four
                equal-area children
                $(c_x \pm \tfrac{\delta}{2},\; c_y \pm \tfrac{\delta}{2},\;
                \tfrac{\delta}{2})$,
                each of which is retained only if it cannot immediately be
                proved unavailable.
        \end{enumerate}
\end{enumerate}

The loop terminates when the voxel list is empty.  At that point every
point in $[0,L]^2$ has been shown to lie outside $\Omega_A \cap \Omega_B$,
or to be covered by the exclusion zone of at least one deposited particle,
up to the residual area $\Delta A \le N_{\mathrm{res}}\,(2\delta_{\min})^2$
from voxels deleted under criterion~(a).  This constitutes a formal
geometric certificate of saturation, not merely a heuristic stopping
criterion.  In all runs reported here the resulting error in packing
fraction satisfies $\Delta\phi < 10^{-6}$.

\subsection{Domain geometry and border clearance}
\label{sec:border}

All polygons are placed with a fixed orientation (first vertex pointing
upward).  For a regular $n$-gon of side length $s$, the circumradius and
inradius are
\begin{equation}
  R_{\mathrm{out}} = \frac{s}{2\sin(\pi/n)},
  \qquad
  R_{\mathrm{in}}  = \frac{s}{2\tan(\pi/n)}.
\end{equation}
Because a fixed-orientation polygon is asymmetric about its centre for odd
$n$, the clearances required at each wall differ between directions.  We
define the \emph{bounding-box extents}
\begin{equation}
  \mathrm{BBOX} = \bigl[b_L,\, b_R,\, b_B,\, b_T\bigr]
  \;\triangleq\;
  \bigl[-x_{\min},\; x_{\max},\; -y_{\min},\; y_{\max}\bigr],
\end{equation}
where $(x_{\min}, x_{\max}, y_{\min}, y_{\max})$ are the extremal vertex
coordinates of the polygon centred at the origin.  A 
implementation used $R_{\mathrm{out}}$ as a uniform clearance in all four
directions, can incorrectly exclue a strip of width
\begin{equation}
  \Delta b_d = R_{\mathrm{out}} - b_d \;\ge\; 0
\end{equation}
near each wall, where $b_d$ is the true extent in direction $d$.  For a
triangle ($n = 3$) this strip has width $R_{\mathrm{out}}/2$ at the bottom
wall and $R_{\mathrm{out}} - R_{\mathrm{in}}$ at the side walls — a
non-negligible fraction of $R_{\mathrm{out}}$.  

\subsection{The \texttt{proveUnavailable} predicate}
\label{sec:prove}

A voxel $V = \{(x,y) : |x - c_x| \le \delta,\; |y - c_y| \le \delta\}$
is declared \emph{unavailable} when it is proved that \emph{neither} species
can be inserted with its centre anywhere in $V$:
\begin{equation}
  \texttt{proveUnavailable}(V) 
 \texttt =
  \texttt{proveSpec}(V,A)
  \;\wedge\;
  \texttt{proveSpec}(V,B).
\end{equation}

\subsubsection{Per-species proof: \texttt{proveSpec}}

\texttt{proveSpec}$(V, \alpha)$ returns \texttt{true} when it can
rigorously demonstrate that species $\alpha$ has no valid centre in $V$.
Three tests are applied in order of increasing computational cost; the
predicate returns as soon as any test succeeds.

\paragraph{Test 1 — Wall exclusion.}
$V$ lies entirely outside $\Omega_\alpha$ if any face of the voxel's
bounding box is fully beyond the corresponding wall clearance:
\begin{equation}
  (c_x + \delta) < b_L^\alpha
  \;\vee\; (c_x - \delta) > L - b_R^\alpha
  \;\vee\; (c_y + \delta) < b_B^\alpha
  \;\vee\; (c_y - \delta) > L - b_T^\alpha.
\end{equation}

\paragraph{Test 2 — Circumradius and inradius bounds.}
For each deposited particle $p$ of species $\beta$ at position
$\mathbf{p} = (p_x, p_y)$, the closest and farthest distances from any
point in $V$ to $\mathbf{p}$ are
\begin{align}
  d_{\min}^2 &= \max(0,\,|p_x - c_x| - \delta)^2
              + \max(0,\,|p_y - c_y| - \delta)^2,\\
  d_{\max}   &= \sqrt{(|p_x - c_x| + \delta)^2
              + (|p_y - c_y| + \delta)^2}.
\end{align}
\begin{itemize}
  \item If $d_{\min} > R_{\mathrm{out}}^\alpha + R_{\mathrm{out}}^\beta$, the
        polygons cannot overlap at any centre in $V$; particle $p$ is skipped.
  \item If $d_{\max} < R_{\mathrm{in}}^\alpha + R_{\mathrm{in}}^\beta$, every
        centre in $V$ places the inscribed circles of the two polygons
        overlapping, so overlap is certain for the entire voxel;
        the voxel is proved unavailable.
\end{itemize}

\paragraph{Test 3 — Robust edge-crossing proof (\texttt{oProve}).}
When neither distance bound is decisive, a geometric certificate is sought.
Two convex polygons intersect if and only if some edge of one crosses an
edge of the other (when neither polygon is fully contained in the other,
which is already handled by Test~2).  \texttt{oProve} searches all edge
pairs $(e_\alpha, e_\beta)$ for a \emph{robust crossing}: one that persists
for every possible centre position within $V$.

\subsection{Robust edge-crossing certificate: \texttt{oProve}}
\label{sec:oprove}

Let the existing particle have fixed vertices $\mathbf{r}_1, \mathbf{r}_2$
(one edge), and let the trial polygon be placed at the voxel centre with
vertices $\mathbf{r}_3, \mathbf{r}_4$ (one edge).  Define the signed
cross-product orientation function
\begin{equation}
  \omega(\mathbf{a}, \mathbf{b}, \mathbf{c})
  = (b_y - a_y)(c_x - b_x) - (b_x - a_x)(c_y - b_y).
  \label{eq:omega}
\end{equation}
The four orientation values for this edge pair are
\begin{equation}
\begin{aligned}
O_1 &= \omega(\mathbf{r}_1,\mathbf{r}_2,\mathbf{r}_3), \qquad
O_2 = \omega(\mathbf{r}_1,\mathbf{r}_2,\mathbf{r}_4), \\
O_3 &= \omega(\mathbf{r}_3,\mathbf{r}_4,\mathbf{r}_1), \qquad
O_4 = \omega(\mathbf{r}_3,\mathbf{r}_4,\mathbf{r}_2).
\end{aligned}
\end{equation}
The two segments cross if and only if $O_1 O_2 < 0$ and $O_3 O_4 < 0$.

When the trial centre shifts by $(\Delta x, \Delta y)$ with
$|\Delta x|, |\Delta y| \le \delta$, the trial vertices $\mathbf{r}_3$,
$\mathbf{r}_4$ each shift by the same amount.  The resulting perturbations
to the orientation values are bounded by
\begin{align}
|\delta O_1|,\; |\delta O_2|
&\le
\bigl(|r_{2y}-r_{1y}|+|r_{2x}-r_{1x}|\bigr)\,\delta
\nonumber\\
&\qquad\triangleq
\|\mathbf{e}_{\mathrm{exist}}\|_1\,\delta ,
\\[6pt]
|\delta O_3|
&\le
2\delta
\Bigl(
|r_{1x}-r_{4x}|+|r_{1y}-r_{4y}|+2\delta
\Bigr)
\nonumber\\
&\qquad
+\delta
\Bigl(
|r_{4y}-r_{3y}|+|r_{4x}-r_{3x}|
\Bigr),
\\[6pt]
|\delta O_4|
&\le
2\delta
\Bigl(
|r_{2x}-r_{4x}|+|r_{2y}-r_{4y}|+2\delta
\Bigr)
\nonumber\\
&\qquad
+\delta
\Bigl(
|r_{4y}-r_{3y}|+|r_{4x}-r_{3x}|
\Bigr).
\end{align}
Here $|\delta O_3|$ and $|\delta O_4|$ are larger than $|\delta O_1|$ and
$|\delta O_2|$ because both endpoints of the trial edge move simultaneously.
An edge crossing is \emph{certified robust} — present for \emph{every}
trial centre in $V$ — when
\begin{equation}
\begin{aligned}
|O_1| &> |\delta O_1|,
&\qquad
|O_2| &> |\delta O_2|,
&\qquad
O_1 O_2 &< 0,
\\
|O_3| &> |\delta O_3|,
&\qquad
|O_4| &> |\delta O_4|,
&\qquad
O_3 O_4 &< 0.
\end{aligned}
\label{eq:robust}
\end{equation}
If conditions~\eqref{eq:robust} hold for at least one edge pair $(e_\alpha,
e_\beta)$, the two polygons must overlap everywhere in $V$, and the voxel
is proved unavailable.

\subsection{Convergence and correctness}
\label{sec:convergence}

The algorithm is \emph{conservative}: \texttt{proveUnavailable} returns
\texttt{true} only when unavailability has been rigorously established, and
never removes a voxel that contains a valid insertion centre.  Voxels for
which the proof fails are subdivided and retried.  As $\delta \to 0$ the
perturbation bounds $|\delta O_k| \to 0$, so the robust-crossing
condition~\eqref{eq:robust} eventually becomes satisfiable for any
genuinely blocked region.  The voxel list therefore shrinks monotonically
and the algorithm terminates in finite time, subject to the $\delta_{\min}$
floor described in Section~\ref{sec:alg-unified}.

\begin{figure}[htbp]
\centering

\begin{subfigure}{0.32\textwidth}
    \centering
    \includegraphics[width=\textwidth]{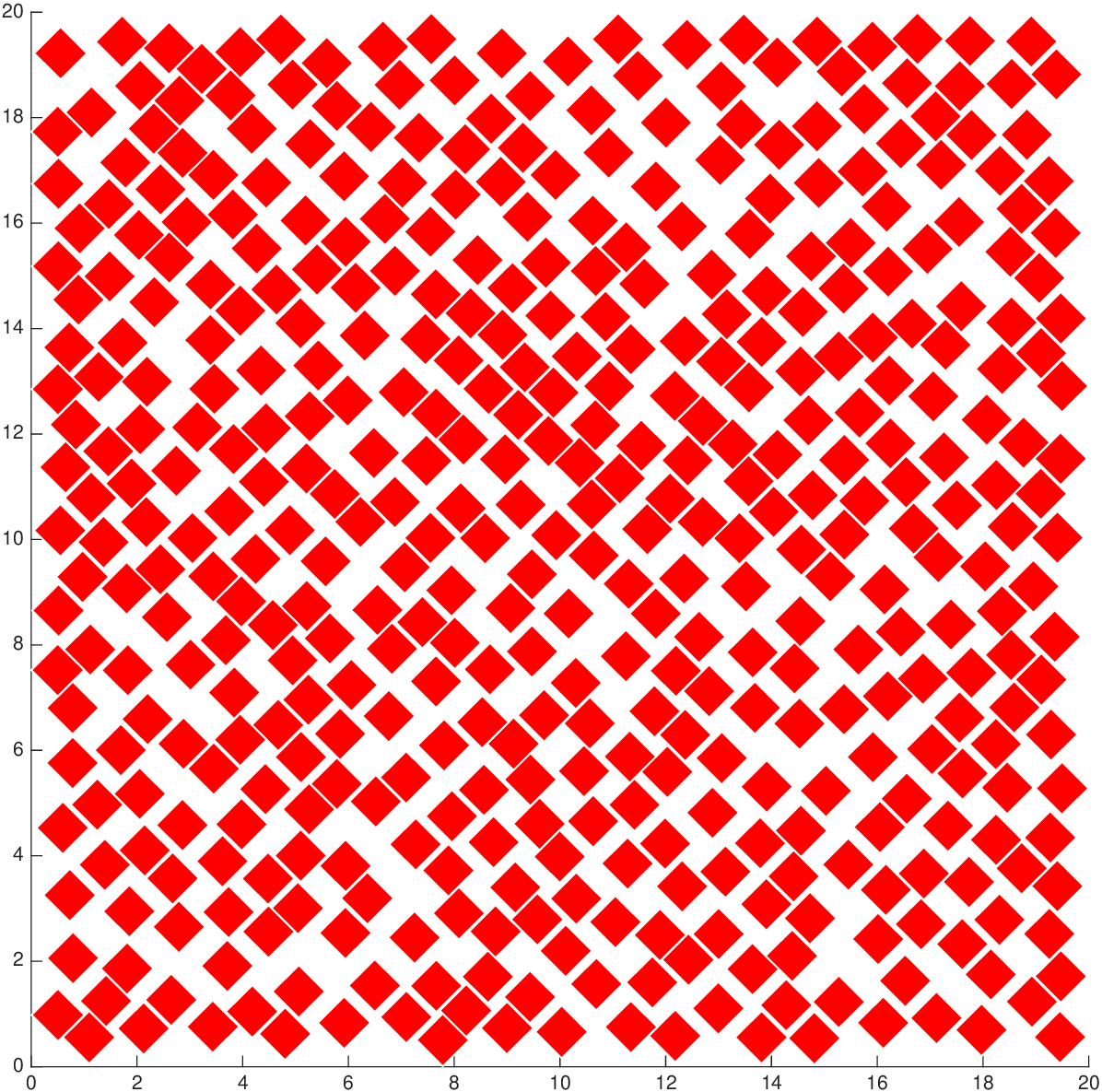}
    \caption{Pure square adsorption}
\end{subfigure}
\hfill
\begin{subfigure}{0.32\textwidth}
    \centering
    \includegraphics[width=\textwidth]{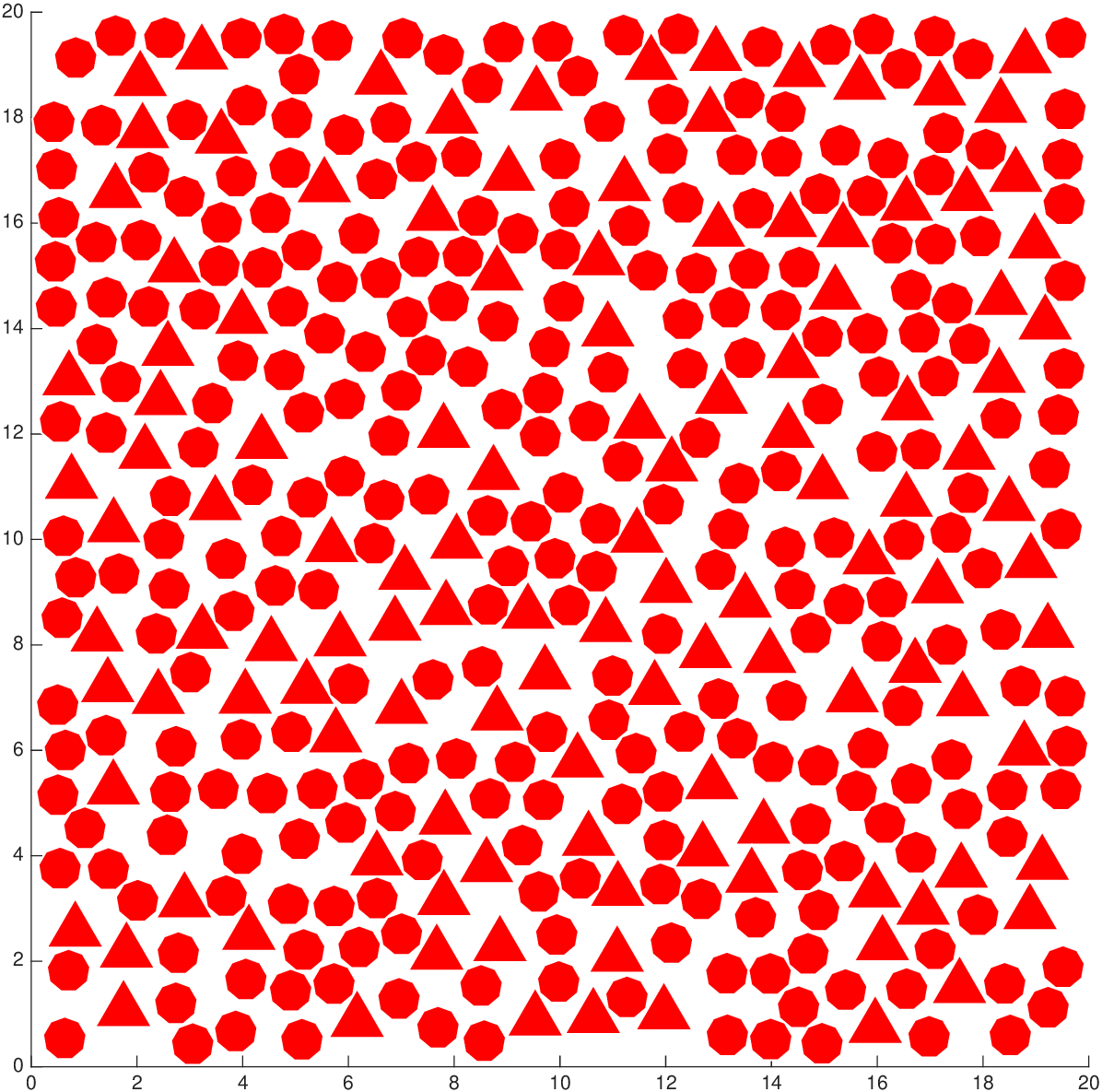}
    \caption{Triangle--nonagon mixture}
\end{subfigure}
\hfill
\begin{subfigure}{0.32\textwidth}
    \centering
    \includegraphics[width=\textwidth]{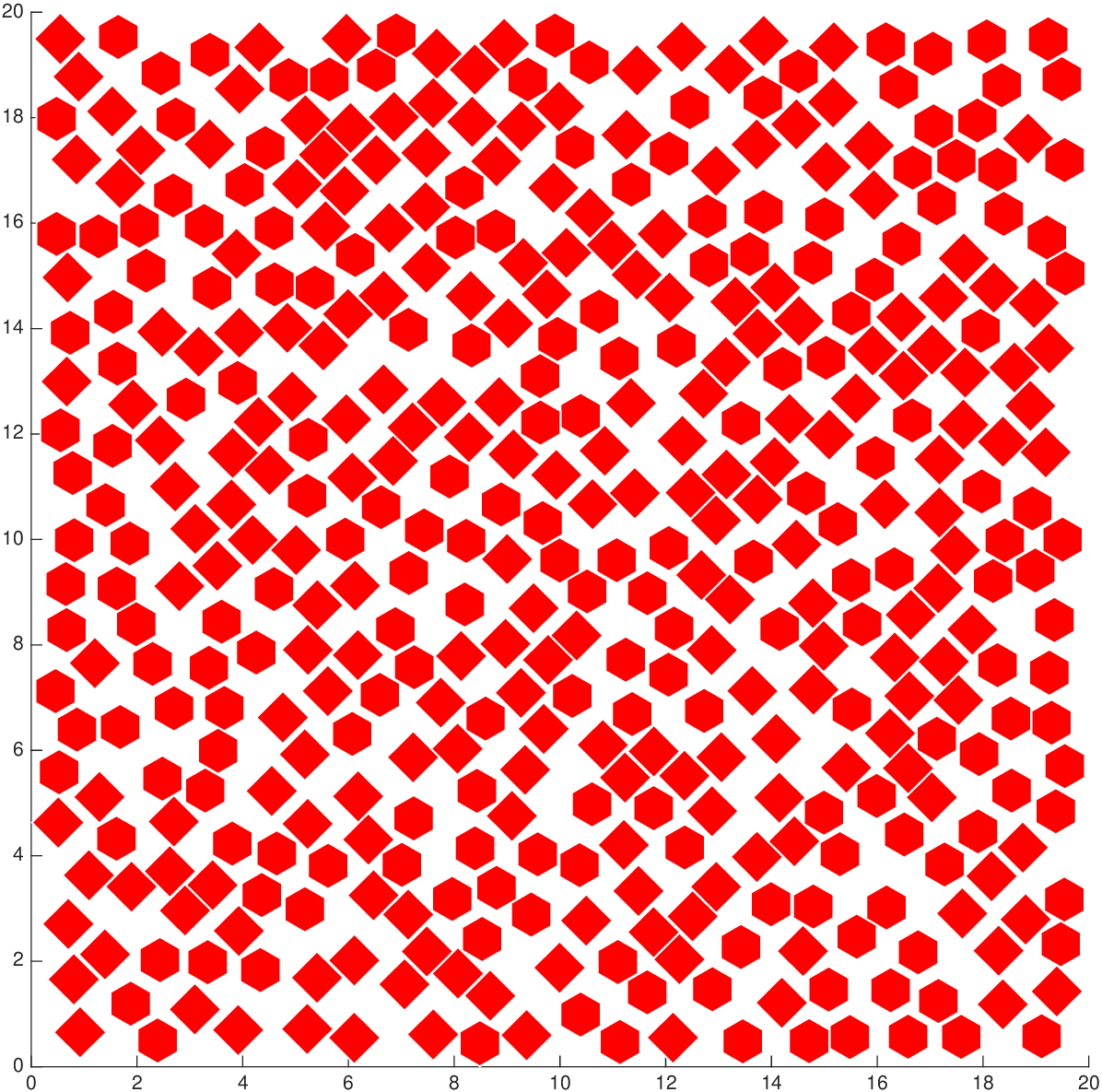}
    \caption{Square--hexagon mixture}
\end{subfigure}

\caption{Representative RSA deposition patterns at saturation for selected regular-polygon systems. Shown are three representative configurations illustrating the influence of particle shape and cross-species geometry on local packing structure and residual void formation. All particles have equal area and fixed orientation.}
\label{fig:rsa_patterns}
\end{figure}

\section{Results and Discussion}

Simulations of several representative jammed configurations are shown in Figure~1; moreover, the complete results of the study are summarized in Table~1. The figure
illustrates the system at the saturation limit for the pure
square case, as well as the jammed configuration obtained
for the triangle--nonagon binary mixture and Square-hexagon binary mixtures as an example.

\begin{table}[ht]
\centering

\caption{Saturation coverage for binary mixtures of equal-area regular polygons.
Values are averaged over independent RSA realizations and reported as
$\phi=\langle\phi\rangle \pm \sigma$, where $\sigma$ is the standard deviation.}

\label{tab:binary_results}

\begin{tabular}{ccccc}
\hline
Combo & Mean $\phi_A$ & Mean $\phi_B$ & Mean Total & Std.\ Dev. \\
\hline

3+3  & 0.183750 & 0.183850 & 0.367600 & 0.002980 \\
3+4  & 0.156225 & 0.364150 & 0.520375 & 0.002724 \\
3+5  & 0.178700 & 0.316650 & 0.495350 & 0.002569 \\
3+6  & 0.156025 & 0.368300 & 0.524325 & 0.002698 \\
3+7  & 0.181625 & 0.335500 & 0.517125 & 0.002652 \\
3+8  & 0.169125 & 0.362375 & 0.531500 & 0.004596 \\
3+9  & 0.175500 & 0.343875 & 0.519375 & 0.002652 \\

4+4  & 0.283233 & 0.281358 & 0.564591 & 0.003594 \\
4+5  & 0.314125 & 0.217813 & 0.531937 & 0.002703 \\
4+6  & 0.285625 & 0.264500 & 0.550125 & 0.001945 \\
4+7  & 0.305875 & 0.241000 & 0.546875 & 0.001591 \\
4+8  & 0.281125 & 0.262250 & 0.543375 & 0.006187 \\
4+9  & 0.293250 & 0.253250 & 0.546500 & 0.001836 \\

5+5  & 0.245700 & 0.245925 & 0.491625 & 0.002829 \\
5+6  & 0.220625 & 0.315200 & 0.535825 & 0.001679 \\
5+7  & 0.237175 & 0.286650 & 0.523825 & 0.003969 \\
5+8  & 0.220375 & 0.316050 & 0.536425 & 0.002342 \\
5+9  & 0.233925 & 0.298975 & 0.532900 & 0.002484 \\

6+6  & 0.273714 & 0.279857 & 0.553571 & 0.002932 \\
6+7  & 0.298083 & 0.242167 & 0.540250 & 0.002385 \\
6+8  & 0.267250 & 0.279583 & 0.546833 & 0.003761 \\
6+9  & 0.288125 & 0.252625 & 0.540750 & 0.001414 \\

7+7  & 0.261917 & 0.258500 & 0.520417 & 0.004446 \\
7+8  & 0.241583 & 0.295917 & 0.537500 & 0.001090 \\
7+9  & 0.258000 & 0.276625 & 0.534625 & 0.001237 \\

8+8  & 0.272583 & 0.277750 & 0.550333 & 0.000764 \\
8+9  & 0.286833 & 0.256333 & 0.543167 & 0.001627 \\
8+10 & 0.273500 & 0.275333 & 0.548833 & 0.002673 \\
8+11 & 0.285167 & 0.260500 & 0.545667 & 0.002097 \\
8+12 & 0.270667 & 0.277083 & 0.547750 & 0.003072 \\
8+13 & 0.269583 & 0.274667 & 0.544250 & 0.001521 \\
8+14 & 0.267833 & 0.278500 & 0.546333 & 0.003166 \\
8+15 & 0.282917 & 0.262583 & 0.545500 & 0.002165 \\
8+16 & 0.271250 & 0.272667 & 0.543917 & 0.004585 \\
8+17 & 0.274417 & 0.273167 & 0.547583 & 0.000144 \\
8+18 & 0.268917 & 0.278250 & 0.547167 & 0.002754 \\
8+19 & 0.276167 & 0.268417 & 0.544583 & 0.003086 \\
8+20 & 0.265833 & 0.283667 & 0.549500 & 0.002222 \\
8+21 & 0.273500 & 0.273000 & 0.546500 & 0.000901 \\
8+22 & 0.273417 & 0.272500 & 0.545917 & 0.002376 \\
8+23 & 0.274083 & 0.272917 & 0.547000 & 0.003041 \\

9+9  & 0.267500 & 0.266875 & 0.534375 & 0.000177 \\

10+10 & 0.268417 & 0.282333 & 0.550750 & 0.001953 \\
11+11 & 0.267167 & 0.270917 & 0.538083 & 0.001665 \\
12+12 & 0.273167 & 0.275833 & 0.549000 & 0.003122 \\
13+13 & 0.256250 & 0.285250 & 0.541500 & 0.001265 \\

14+14 & 0.272725 & 0.275200 & 0.547925 & 0.002430 \\
14+15 & 0.285850 & 0.262175 & 0.548025 & 0.001766 \\
14+16 & 0.273675 & 0.273225 & 0.546900 & 0.002824 \\
14+17 & 0.279028 & 0.267667 & 0.546694 & 0.003295 \\
16+16 & 0.273900 & 0.275800 & 0.549700 & 0.002900 \\
17+17 & 0.270375 & 0.273688 & 0.544063 & 0.003716 \\

\hline
\end{tabular}

\end{table}

Table~1 summarizes saturation coverages for mixtures spanning from pure triangular systems to higher-order binary combinations. Triangle-containing mixtures consistently yield lower saturation values than systems composed of larger polygons. In these mixtures, a systematic asymmetry emerges, with higher-order polygons contributing more strongly to the total coverage. The highest saturation coverage, is obtained for the pure square system. In particular, the aligned square--square configuration yields the maximum measured coverage across all studied mixtures.

\begin{equation}
\phi_{\mathrm{sat}}\approx0.5646.
\end{equation}

This value slightly exceeds the classical RSA saturation value for disks and represents the maximum among all mixtures investigated in the present study.

As the polygon order increases,

\begin{equation}
n\rightarrow\infty,
\end{equation}

regular polygons converge toward circular geometry \cite{Ciesla2023}. Accordingly, the saturation coverage is expected to approach the classical disk RSA value reported in the literature.

All simulations used symmetric deposition conditions ($p_A=p_B=\tfrac{1}{2}$),
yet the jammed states exhibit strong composition-dependent partitioning of
coverage between species.

\subsection{Excluded-Area Analysis}
\label{sec:excluded_area}

A direct geometric explanation for the saturation-coverage trends observed in
Table~\ref{tab:binary_results} is provided by the \emph{excluded area}
$E_{AB}$—the region within which the centre of a particle of species~$B$
cannot be placed once a particle of species~$A$ has already been deposited.
For two fixed-orientation convex polygons $P_A$ and $P_B$,
\begin{equation}
  E_{AB} \;=\; \mathrm{Area}\!\bigl(P_A \oplus (-P_B)\bigr),
  \label{eq:excluded_def}
\end{equation}
where $\oplus$ denotes the Minkowski sum and $-P_B$ is the reflection of
$P_B$ through the origin.  For fixed-orientation regular polygons this
quantity is analytically tractable: the Minkowski sum of two convex polygons
with $n_A$ and $n_B$ sides is itself a convex polygon, and its area can be
computed exactly by merging the sorted edge-vector sequences of $P_A$ and
$-P_B$~\cite{de2000computational}.

\subsubsection*{Analytical result for same-species pairs}

For the \emph{same-species} excluded area $E_{AA} = \mathrm{Area}(P_A \oplus
(-P_A))$ an exact closed-form expression follows from centrosymmetry
arguments:

\begin{itemize}
  \item \textbf{Even $n$:} A regular $n$-gon with $n$ even is centrosymmetric,
    i.e.\ $-P = P$.  Therefore $P_A \oplus (-P_A) = P_A \oplus P_A = 2P_A$,
    and
    \begin{equation}
      E_{AA}^{(n\;\text{even})} \;=\; \mathrm{Area}(2P_A) \;=\; 4A_0,
      \label{eq:E_even}
    \end{equation}
    independently of $n$.

  \item \textbf{Odd $n$:} A regular $n$-gon with $n$ odd is \emph{not}
    centrosymmetric, so $-P_A \neq P_A$ and the Minkowski sum is strictly
    larger than $2P_A$.  Explicit vertex enumeration gives
    \begin{equation}
      E_{AA}^{(n\;\text{odd})} \;=\; 4A_0\!\left[1 + \tfrac{1}{2}
        \!\left(\sec\tfrac{\pi}{n} - 1\right)\right]
      \;=\; 4A_0 + 2A_0\!\left(\sec\tfrac{\pi}{n} - 1\right),
      \label{eq:E_odd}
    \end{equation}
    which decreases monotonically toward $4A_0$ as $n\to\infty$ (approaching
    the centrosymmetric circle limit).
\end{itemize}

Numerical values computed via the Minkowski-sum algorithm are in exact
agreement with Eqs.~\eqref{eq:E_even}--\eqref{eq:E_odd}; see
Figure~\ref{fig:E_vs_n}.  The even/odd parity of $n$ therefore produces a
systematic alternation in same-species excluded area that directly mirrors the
alternation in saturation coverage visible in
Table~\ref{tab:binary_results}: even-sided polygons pack more efficiently
because their reflected copies fit together with the same efficiency as the
originals.

\subsubsection*{Cross-species excluded area and correlation with $\phi_{\rm sat}$}

\begin{figure*}[htbp]
  \centering
  \includegraphics[width=0.95\textwidth]{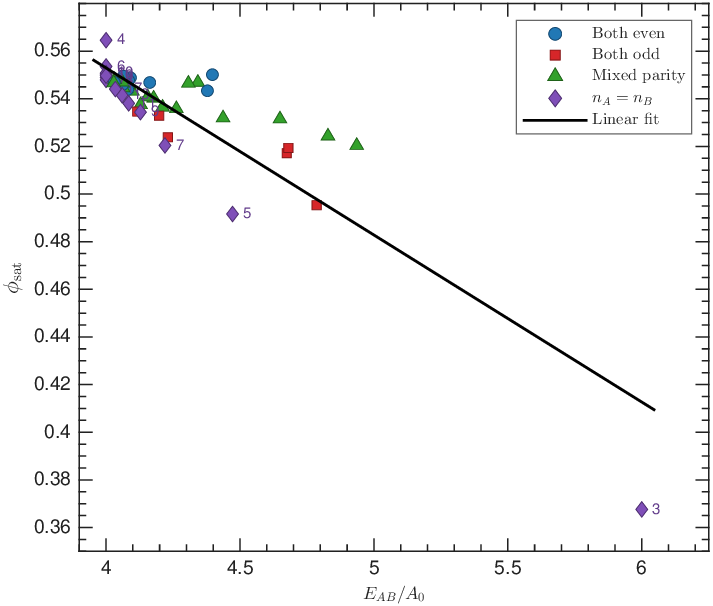}
 \caption{Saturation coverage $\phi_{\rm sat}$ as a function of the
    normalised cross-species excluded area $E_{AB}/A_0$.
    Marker shape and colour encode parity: both-even pairs (circles, blue),
    both-odd pairs (squares, red), mixed-parity pairs (triangles, green), and
    same-species diagonal ($n_A=n_B$, diamonds, purple).
    The solid line is a least-squares linear fit to all points.
    Both-even pairs lie systematically above the trend, confirming that the
    centrosymmetry of even-sided polygons ($E_{AA}=4A_0$) is the primary
    driver of their enhanced packing.}
  \label{fig:phi_vs_E}
\end{figure*}

\begin{figure*}[htbp]
  \centering
  \includegraphics[width=0.95\textwidth]{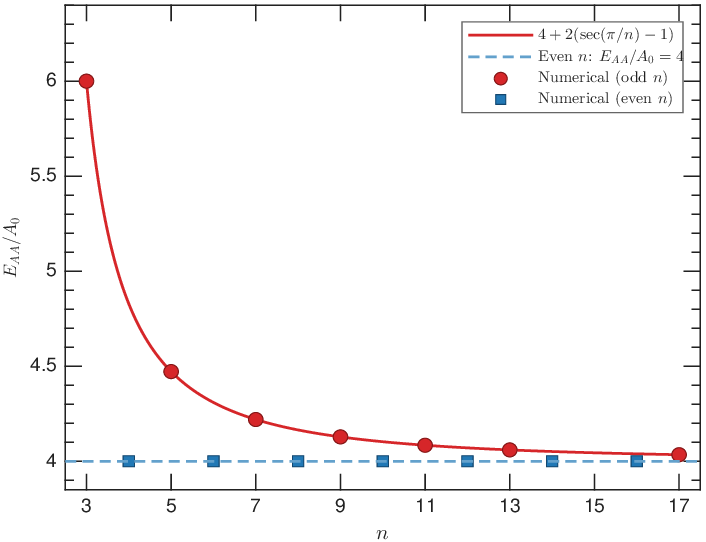}
  \caption{Same-species normalised excluded area $E_{AA}/A_0$ as a function of
    polygon side count $n$.  Even $n$ achieves the minimum value of exactly
    $4A_0$ (horizontal dashed line) due to centrosymmetry.  Odd $n$ follows
    the analytical formula $E_{AA}/A_0 = 4 + 2(\sec(\pi/n)-1)$
    (solid curve), converging to $4A_0$ from above as $n\to\infty$.}
  \label{fig:E_vs_n}
\end{figure*}

Table~\ref{tab:excluded_area} lists $E_{AB}/A_0$ for every pair together with
the measured saturation coverage.  Figure~\ref{fig:phi_vs_E} shows that
$\phi_{\rm sat}$ is \emph{negatively} correlated with $E_{AB}/A_0$, following standard mean-field RSA arguments, one may heuristically estimate 
\begin{equation}
  \phi_{\rm sat} \;\approx\; \frac{A_0}{\tilde{E}}\,
    \left(1 - e^{-\phi_{\rm sat}\,\tilde{E}/A_0}\right)^{-1},
  \label{eq:mf_rsa}
\end{equation}
where $\tilde{E} = (E_{AA}+2E_{AB}+E_{BB})/4$ is an effective excluded area
averaged over the two species (each chosen with equal probability).  Pairs in
which \emph{both} species have an even number of sides (blue circles in
Figure~\ref{fig:phi_vs_E}) systematically lie \emph{above} the global trend
line because their same-species excluded area achieves the minimum value
$4A_0$, giving the mixture the most room to accommodate additional particles
before jamming.

\begin{table}[ht]
\centering
\caption{Normalised excluded areas and saturation coverage for all binary
  pairs.  $E_{AA}$ and $E_{BB}$ are same-species excluded areas;
  $E_{AB}=\mathrm{Area}(P_A\oplus(-P_B))$ is the cross-species excluded area.
  All values normalised by $A_0=0.1$.  The $\phi_{\rm sat}$ column reproduces
  the mean total coverage from Table~\ref{tab:binary_results} exactly.}
\label{tab:excluded_area}
\small
\begin{tabular}{c ccc c}
\toprule
Combo & $E_{AA}/A_0$ & $E_{BB}/A_0$ & $E_{AB}/A_0$ & $\phi_{\rm sat}$ \\
\midrule

3+3  & 6.0000 & 6.0000 & 6.0000 & 0.367600 \\
3+4  & 6.0000 & 4.0000 & 4.9358 & 0.520375 \\
3+5  & 6.0000 & 4.4721 & 4.7862 & 0.495350 \\
3+6  & 6.0000 & 4.0000 & 4.8284 & 0.524325 \\
3+7  & 6.0000 & 4.2198 & 4.6744 & 0.517125 \\
3+8  & 6.0000 & 4.0000 & 4.6492 & 0.531500 \\
3+9  & 6.0000 & 4.1284 & 4.6806 & 0.519375 \\

4+4  & 4.0000 & 4.0000 & 4.0000 & 0.564591 \\
4+5  & 4.0000 & 4.4721 & 4.4368 & 0.531937 \\
4+6  & 4.0000 & 4.0000 & 4.3971 & 0.550125 \\
4+7  & 4.0000 & 4.2198 & 4.3426 & 0.546875 \\
4+8  & 4.0000 & 4.0000 & 4.3784 & 0.543375 \\
4+9  & 4.0000 & 4.1284 & 4.3074 & 0.546500 \\

5+5  & 4.4721 & 4.4721 & 4.4721 & 0.491625 \\
5+6  & 4.4721 & 4.0000 & 4.2625 & 0.535825 \\
5+7  & 4.4721 & 4.2198 & 4.2308 & 0.523825 \\
5+8  & 4.4721 & 4.0000 & 4.2111 & 0.536425 \\
5+9  & 4.4721 & 4.1284 & 4.1979 & 0.532900 \\

6+6  & 4.0000 & 4.0000 & 4.0000 & 0.553571 \\
6+7  & 4.0000 & 4.2198 & 4.1775 & 0.540250 \\
6+8  & 4.0000 & 4.0000 & 4.1631 & 0.546833 \\
6+9  & 4.0000 & 4.1284 & 4.1554 & 0.540750 \\

7+7  & 4.2198 & 4.2198 & 4.2198 & 0.520417 \\
7+8  & 4.2198 & 4.0000 & 4.1289 & 0.537500 \\
7+9  & 4.2198 & 4.1284 & 4.1164 & 0.534625 \\

8+8  & 4.0000 & 4.0000 & 4.0000 & 0.550333 \\
8+9  & 4.0000 & 4.1284 & 4.0981 & 0.543167 \\
8+10 & 4.0000 & 4.0000 & 4.0911 & 0.548833 \\
8+11 & 4.0000 & 4.0844 & 4.0832 & 0.545667 \\
8+12 & 4.0000 & 4.0000 & 4.0840 & 0.547750 \\
8+13 & 4.0000 & 4.0599 & 4.0747 & 0.544250 \\
8+14 & 4.0000 & 4.0000 & 4.0727 & 0.546333 \\
8+15 & 4.0000 & 4.0447 & 4.0695 & 0.545500 \\
8+16 & 4.0000 & 4.0000 & 4.0808 & 0.543917 \\
8+17 & 4.0000 & 4.0346 & 4.0661 & 0.547583 \\
8+18 & 4.0000 & 4.0000 & 4.0652 & 0.547167 \\
8+19 & 4.0000 & 4.0277 & 4.0636 & 0.544583 \\
8+20 & 4.0000 & 4.0000 & 4.0647 & 0.549500 \\
8+21 & 4.0000 & 4.0226 & 4.0619 & 0.546500 \\
8+22 & 4.0000 & 4.0000 & 4.0615 & 0.545917 \\
8+23 & 4.0000 & 4.0188 & 4.0606 & 0.547000 \\

9+9  & 4.1284 & 4.1284 & 4.1284 & 0.534375 \\

10+10 & 4.0000 & 4.0000 & 4.0000 & 0.550750 \\
11+11 & 4.0844 & 4.0844 & 4.0844 & 0.538083 \\
12+12 & 4.0000 & 4.0000 & 4.0000 & 0.549000 \\
13+13 & 4.0599 & 4.0599 & 4.0599 & 0.541500 \\

14+14 & 4.0000 & 4.0000 & 4.0000 & 0.547925 \\
14+15 & 4.0000 & 4.0447 & 4.0321 & 0.548025 \\
14+16 & 4.0000 & 4.0000 & 4.0304 & 0.546900 \\
14+17 & 4.0000 & 4.0346 & 4.0287 & 0.546694 \\

16+16 & 4.0000 & 4.0000 & 4.0000 & 0.549700 \\
17+17 & 4.0346 & 4.0346 & 4.0346 & 0.544063 \\

\bottomrule
\end{tabular}
\end{table}

\subsection{Convergence of pure-species coverage to the disk RSA limit}
\label{sec:convergence}

For the diagonal entries of the mixture table ($n_A = n_B \equiv n$, pure
single-species systems), the saturation coverage $\phi_{\rm sat}(n)$ is
expected to approach the classical disk RSA limit
$\phi_{\rm disk} \approx 0.547$~\cite{Feder1980} as $n\to\infty$.
Previous work by Cie\'sla \textit{et al.}~\cite{Ciesla2023}
motivates quantifying this convergence rate
\begin{equation}
  \phi_{\rm sat}(n) = \phi_{\rm disk} - \frac{c}{n^{\alpha}},
  \label{eq:convergence_ansatz}
\end{equation}
where the $1/n^2$ exponent ($\alpha=2$) arises naturally because the
isoperimetric deficit of a regular $n$-gon,
\begin{equation}
  \mathcal{D}(n) = 1 - \frac{4\pi A_0}{P_n^2} \;\propto\; \frac{1}{n^2},
\end{equation}
measures the deviation of the polygon boundary from circularity at fixed area.

However, the diagonal data reveal that Eq.~\eqref{eq:convergence_ansatz} cannot
be applied uniformly across all $n$, because the even/odd parity alternation
identified in Section~\ref{sec:excluded_area} produces \emph{non-monotone}
convergence.  Even-sided polygons ($n=4,6,8,10,12,14,16$), whose same-species
excluded area achieves the centrosymmetric minimum $E_{AA}=4A_0$, yield
saturation coverages that \emph{exceed} $\phi_{\rm disk}$:
\begin{equation}
  \phi_{\rm sat}(4) \approx 0.565,\quad
  \phi_{\rm sat}(6) \approx 0.554,\quad
  \phi_{\rm sat}(8) \approx 0.550,
\end{equation}
all above $\phi_{\rm disk} \approx 0.547$, with the higher even-$n$ values
($n=10,12,14,16$) remaining above the disk limit but approaching it from above
as $n$ increases.  These even-$n$ systems therefore converge to the disk limit
\emph{from above}, so the sign of $c$ in Eq.~\eqref{eq:convergence_ansatz}
is negative for even parity.  By contrast, odd-sided polygons
($n=5,7,9,11,13,17$) lie below $\phi_{\rm disk}$ and converge from below.

\subsubsection*{Odd sub-sequence fit}

Restricting the fit to $n \in \{5,7,9,11,13,17\}$ (excluding the triangle, which
is an outlier due to its anomalously large excluded area
$E_{AA}/A_0 = 6.0$) yields
\begin{equation}
  \phi_{\rm sat}(n)\big|_{\rm odd}
  = 0.547 - \frac{2.70 \pm 0.37}{n^{2.41 \pm 0.06}},
  \label{eq:fit_odd}
\end{equation}
with $R^2 = 0.998$.  The fitted exponent $\alpha \approx 2.41$ is
close to, but distinguishably above, the $1/n^2$ prediction, and is
consistent across the full odd sub-sequence up to $n=23$.  This modest
departure from $\alpha = 2$ may reflect the fact that for fixed-orientation
RSA, orientation-dependent packing effects contribute corrections beyond the
purely isoperimetric scaling.

\subsubsection*{Even sub-sequence}

For even $n$, the excess coverage above $\phi_{\rm disk}$ decays as $n$
increases and the polygon approaches a circle.  Fitting to all available
even diagonal points $n\in\{4,6,8,10,12,14,16\}$,
\begin{equation}
  \Delta\phi(n)\big|_{\rm even}
  = \phi_{\rm sat}(n) - \phi_{\rm disk}
  \approx \frac{0.15}{n^{1.69}},
\end{equation}
indicating that the tiling advantage of centrosymmetric shapes vanishes
as a power law in $n$, consistent with the isoperimetric argument.

\subsubsection*{Unified description}

The two sub-sequence behaviours are captured by the parity-corrected form
\begin{equation}
  \phi_{\rm sat}(n)
  = \phi_{\rm disk}
  - \frac{c}{n^{\alpha}}
  + \delta\,(-1)^{n},
  \label{eq:convergence_parity}
\end{equation}
where the term $\delta(-1)^n$ encodes the even/odd centrosymmetry
alternation.  A least-squares fit to all diagonal points ($n=3,\ldots,23$)
yields $\delta = 0.019 \pm 0.008$, consistent with the excluded-area
difference $\Delta E/A_0 = E_{AA}^{\rm odd} - 4A_0 = 2(\sec(\pi/n)-1)$
evaluated at representative $n$, and confirming that the parity term
has the same geometric origin as the coverage alternation reported in
Table~\ref{tab:binary_results}.

These results are broadly consistent with Cie{\'s}la \textit{et al.}~\cite{Ciesla2023},
who observed a similar dependence of $\phi_{\rm sat}$ on $n$ for single-species
oriented-polygon RSA, and extend that observation by identifying the
even/odd bifurcation as the dominant feature of the convergence landscape
across the full range $n=3,\ldots,23$.

Additional notes: 
 
\paragraph{Symmetric pairs.}
For same-species mixtures ($n_A=n_B$), both species contribute nearly equally
to the final coverage, as expected by symmetry.  The $3+3$ system gives
$\phi_A=0.184\pm0.007$ and $\phi_B=0.184\pm0.008$, with
$\phi_{\rm sat}=0.368\pm0.003$.  The same near-equal split ($\phi_A/\phi_{\rm
sat}\approx50\%$) is recovered for $6+6$, $7+7$, $8+8$, and $9+9$,
confirming that coverage asymmetries observed in other pairs are emergent
effects of geometric incompatibility rather than sampling bias.
 
\paragraph{Even-odd asymmetry.}
When even- and odd-sided polygons are paired, the even-sided species
systematically captures the larger share of the jammed coverage.  This behavior is evident in mixtures such as $4+5$, $4+7$, $6+7$, $6+9$, and
$8+9$. The even-sided species accounts for approximately
$\phi_A/\phi_{\rm sat}\approx59\%$ of the saturated coverage in the $4+5$
mixture, $55\%$ in $6+7$, and $54\%$ in $8+9$. This preferential occupation
originates from the centrosymmetry of even-sided polygons, which guarantees
the minimum same-species excluded area, $E_{AA}=4A_0$. Consequently, these
particles experience less geometric self-exclusion and are able to exploit
the available deposition space more effectively than their odd-sided
counterparts.
 
\paragraph{Triangle-containing mixtures.}
Mixtures involving the triangular species (such as $3+4$, $3+6$, $3+8$) show the
strongest partitioning asymmetry: the triangle fraction contributes only
$\phi_A/\phi_{\rm sat}\approx30\%$, with the competing species absorbing the
remainder.  The anomalously large triangle excluded area ($E_{AA}/A_0=6.0$,
compared to the even-polygon minimum of $4.0$) restricts the admissible
configuration space for triangles disproportionately, driving a substantial
redistribution of surface coverage toward the higher-order species.
 
\paragraph{Global jamming constraint.}
Despite these large shifts in partial coverages, the total coverage
$\phi_{\rm tot}$ is remarkably stable across all mixtures, varying weakly
around $0.54\pm0.02$ for all pairs with $n\ge6$.  This stability indicates
a global jamming constraint that is largely insensitive to species composition
but strongly sensitive to the effective excluded area $\tilde{E}$.  Partial
coverage fluctuations therefore reflect a redistribution of surface occupancy
between species rather than any change in overall packing efficiency.
 
\paragraph{Convergence to the disk limit.}
For higher-order polygon pairs ($n\ge6$), saturation coverages cluster within
$\phi_{\rm sat}\approx0.53$--$0.55$, converging toward the classical disk RSA
value $\phi_{\rm disk}\approx0.547$~\cite{Feder1980}.  As   quantified in
Section~\ref{sec:convergence}, this convergence is non-monotone: even-$n$
species approach $\phi_{\rm disk}$ from above while odd-$n$ species approach
it from below, with the odd sub-sequence following
$\phi_{\rm sat}(n)=\phi_{\rm disk}-c/n^{\alpha}$ with $\alpha\approx2.41$,
close to the $1/n^2$ isoperimetric prediction.

\section{Conclusions}
 
We have investigated binary RSA mixtures of equal-area regular polygons
using an adaptive split-voxel algorithm with exact overlap detection via
the Separating Axis Theorem. The framework yields accurate saturation
coverages across all 52 unique pairs with $n\in\{3,\dots,23\}$ and
establishes a complete shape-space map for fixed-orientation binary
deposition.
 
The results are governed by two competing geometric mechanisms: gap-filling
enhancement, which benefits combinations involving squares and higher-order
polygons, and cross-species exclusion suppression, which dominates
triangle-containing systems and sharply reduces accessible configuration
space. These effects are captured directly through the Minkowski-sum excluded
area $E_{AB}$: saturation coverage is negatively correlated with $E_{AB}$,
centrosymmetric even-$n$ polygons achieve the minimum same-species excluded
area $E_{AA}=4A_0$, and the parity alternation in $\phi_{\rm sat}$ follows
directly from this geometric distinction. For the odd-$n$ sub-sequence,
pure-species coverage converges to the disk RSA limit as
$\phi_{\rm sat}(n)=\phi_{\rm disk}-c/n^{\alpha}$ with $\alpha\approx2.41$,
close to the $1/n^2$ isoperimetric prediction, while even-$n$ systems
approach $\phi_{\rm disk}$ from above, confirming their persistent tiling
advantage. These findings are consistent with and extend the single-species
results of Cie{\'s}la \textit{et al.}~\cite{Ciesla2023}.
 
The principal findings are: (i)~triangle-containing mixtures strongly
suppress saturation; (ii)~axis-aligned squares yield the maximum measured
coverage $\phi_{\rm sat}\approx0.5646$; (iii)~even-sided polygons hold a
systematic parity-driven adsorption advantage; and (iv)~coverage converges
to the disk limit with an exponent $\alpha\approx2.41$ for odd-$n$ species.
 
The split-voxel framework extends naturally to freely rotating polygons,
non-convex particles, polydisperse mixtures, and three-dimensional
polyhedral RSA, providing a foundation for future theoretical and
experimental studies of shape-driven jamming in multicomponent systems.

\bibliographystyle{apsrev4-2}
\bibliography{refs}

\end{document}